\documentclass{elsart}

\usepackage{amssymb,amsmath}
\usepackage{latexsym}
\usepackage{amsmath}
\usepackage{graphicx}
\journal{Journal of Mechanics and Physics of Solids}
\date{4-Nov.-2010}
\begin{document}

\begin{frontmatter}


\title{Overall Dynamic Constitutive Relations of Micro-structured Elastic Composites}

\author{Sia Nemat-Nasser},
\author{Ankit Srivastava\corauthref{cor1}}
\ead{ansrivas@ucsd.edu}
\corauth[cor1]{Corresponding author.}
\address{Department of Mechanical and Aerospace Engineering\\ University of California, San Diego, La Jolla, CA, 92093-0416\\ USA June 2010}
%\title{}


%\author{}
%\author{Sia Nemat-Nasser, Ankit Srivastava}

%\address{...}

%\address{Department of Mechanical and Aerospace Engineering\\ University of California, San Diego, La Jolla, CA, 92093-0416\\ USA June 2010}

%THIS VERSION WAS LAST MODIFIED 09/19/10

\begin{abstract}
A method for homogenization of a heterogeneous (finite or periodic) elastic composite is presented. 
It allows direct, consistent, and accurate evaluation of the averaged overall frequency-dependent dynamic material constitutive relations.  It is shown that when the spatial variation of the field variables is restricted by a Bloch-form (Floquet-form) periodicity,  then these relations together with the overall conservation and kinematical equations accurately yield the displacement or stress mode-shapes and, necessarily, the dispersion relations.  It also gives as a matter of course point-wise solution of the elasto-dynamic field equations, to any desired degree of accuracy. The resulting overall dynamic constitutive relations however, are general and need not be restricted by the Bloch-form periodicity. The formulation is based on micro-mechanical modeling of a representative unit cell of the composite proposed  by Nemat-Nasser and coworkers; see, e.g., \cite{SNNH} and \cite{SNN_Iwa1}.

We show that, for a micro-structured elastic composite, the overall effective mass-density and compliance (stiffness) are always real-valued and positive, whether or not the corresponding unit cell (representative volume element used as a unit cell) is geometrically and/or materially symmetric. The average strain and linear momentum are however couple and the coupling constitutive parameters are always each others complex conjugates for any  heterogeneous elastic unit cell, such that the overall energy-density is always real and positive.  In this paper, we have sought to separate the overall constitutive relations which should depend only on the composition and structure of the unit cell, from the overall field equations which should hold for any elastic composite; i.e., we use only the \emph{local} field equations and material properties to deduce the \emph{overall} constitutive relations.  

It is shown, by way of an example of a bi-layered composite, that dispersion curves obtained by our method accurately produce the exact results of Rytov \cite{Rytov}. The method is also used to calculate the effective parameters for a 2-layered composite and the results are compared with those of homogenization based on the field integration of the exact solution (Willis \cite{Willis1}, and Nemat-Nasser et al. \cite{NWSA}), and certain relevant issues are clarified.  Finally the method is used to homogenize both a symmetric and a non-symmetric 4-layered composite and the results for the symmetric case are compared with those reported by Nemat-Nasser et al. \cite{NWSA} as well as the exact solution. Thus, this method provides a powerful solution and homogenization tool to use in many cases where the unit cell contains inclusions of complex geometry.
\end{abstract}

\begin{keyword}
Homogenization \sep Microstructured Elastic Composites 
\PACS *43.20.Gp \sep *43.20.Jr \sep 62.20.D-
\end{keyword}
\end{frontmatter}

% main text
\section{Introduction}

The recent interest in the characterization of the overall dynamic properties of composites with tailored microstructure, necessitates a systematic homogenization procedure to express the  dynamic response of an elastic composite in terms of its average effective compliance and  density. The resulting homogenized parameters, once restricted by a Bloch-form periodicity, must give the composite's dispersion relations as well as the corresponding strain or stress mode-shapes.

In the present paper, we propose a method for homogenization of microstructured elastic composites that directly provides the overall frequency-dependent dynamic material parameters. In addition, when the spatial variation of the overall field quantities is restricted to follow a Bloch-type periodicity, then the overall conservation and kinematical relations yield the corresponding stress and displacement mode-shapes and dispersion relations.  The method does not require the values of the field variables within the unit cell. Indeed, it actually gives as a matter of course the point-wise solution of the elasto-dynamic field equations to any desired degree of accuracy. The method is inspired by the micro-mechanical homogenization of the static properties of a typical unit cell of a composite, originally proposed by Nemat-Nasser and Taya \cite{SNN_Taya}, and further developed in \cite{SNN_Iwa_H} and \cite{SNN_Iwa1}.  A similar homogenization method has been used by Amirkhizi and Nemat-Nasser \cite{Ali2} to calculate the effective electromagnetic properties of a periodic composite.  
Here however, we clearly distinguish the overall constitutive relations which should reflect the composition of the unit cell, from the overall conservation and kinematical relations which should hold for any (homogenized) elastic composite with any unit cell. 
In other words, while we do use the \emph{local} (at each point within the unit cell) material properties and do enforce the conservation laws at the local level to obtain the \emph{overall} constitutive relations, the \emph{overall} conservation and kinematical relations are \emph{not} used to derive these overall constitutive relations.   Therefore, the overall constitutive relations can be used to solve initial-boundary value problems for waves of wavelengths that are large relative to the micro-structural length scale.
 
Homogenization for calculating the overall properties of composites, based on the integration of the field variables, has been proposed by a number of researchers. For electromagnetic waves, see, for example, Pendry and Smith \cite{Pendry3,Pendry1,Pendry2}, Amirkhizi and Nemat-Nasser \cite{Ali1}, Bensoussan \cite{Bensoussan}, and Sihvola \cite{Sihvola}. For elasto-dynamic waves, Willis \cite{Willis1} has presented a homogenization method based on an ensemble averaging technique of the 'Bloch' reduced form of the wave propagating in a periodic composite. An analogous approach for the calculation of effective compliance and mass density of a layered composite was worked out in some details by Nemat-Nasser et. al. \cite{NWSA} who showed that the corresponding results yield the exact dispersion relations. 

In what follows, we outline our approach in terms of a layered composite.  We compare, by way of an example of a 2-layered composite, the corresponding dispersion results with those obtained using the exact solution. Their close correspondence indicates that our micro-structurally-based method may be used to calculate the dispersion relation for more complex 2- and 3-dimensional cases, where exact solutions are not available, as we shall report elsewhere.  Here we also use our approach to homogenize a 4-layered composite and compare the results with those obtained by the field integration of the stress and displacement mode-shapes obtained using the mixed variational formulation of Nemat-Nasser et al. \cite{SNN1,SNN2,SNN3}. It is shown that the micro-structurally-based method gives homogenized results which converge to the field integration-based homogenization results that are based on either the exact solution or the mixed variational formulation. Thus, the micro-structurally-based method can be used to evaluate the effective dynamic parameters of complex unit cells, without \emph{a priori} requiring explicit point-wise values of the field variables, but instead producing these quantities if desired.

\section{Micro-structural homogenization of layered composites}

Here we present a homogenization method based on a micro-mechanical consideration of the volume averages of the field variables, viewed as measurable macroscopic physical quantities. We express the solution to the elasto-dynamic equations of motion as the sum of the volume average and a perturbation due to the heterogeneous composition of the unite cell,

\begin{equation}
{Q}={Q}^0+{Q}^p
\end{equation}

where ${Q}$ represents any of the field variables, stress, ${\sigma}$, or velocity, $\dot{u}$. 
The aim is to derive a set of constitutive relations for the overall averaged parts of the field variables, using the local elasto-dynamic equations of motion and constitutive relations.  
This then provides the homogenized frequency-dependent material parameters.  In what follows, we describe our approach using a periodic composite, but the final constitutive relations also apply  to a finite unit cell.

Consider harmonic waves in an unbounded elastic composite consisting of a collection of bonded, 
identical unit cells, 
$\Omega=\{x:-a/2\leq x<a/2\}$, 
which are repeated in the $x$-direction, and hence constitute a periodic structure. 
In view of the periodicity of the composite, we have 
$\rho(x)=\rho(x+m^{'}a)$ and 
$C(x)=C(x+m^{'}a)$; 
here $m^{'}$ is an integer, $\rho(x)$ is the density and 
$C(x)$ is the modulus of elasticity. 
For time harmonic waves with frequency 
$\omega$, the field quantities are proportional to 
$e^{\pm i\omega t}$. 
For waves with arbitrary wavenumber $q$ and unrelated frequency $\omega$, we express the field variables in the following form:

\begin{equation}\label{EBloch}
{\hat{F}(x,t)}=\mathrm{Re}\left[F(x)\exp[i(qx-\omega t)]\right]
\end{equation}

where $\hat{F}$ represents the field variables, stress, 
$\hat{\sigma}$, strain, 
$\hat{\varepsilon}$, momentum, 
$\hat{p}$, displacement, 
$\hat{u}$, or velocity, 
$\hat{\dot{u}}$, 
whereas $F$ represents their periodic parts 
(${\sigma}$, ${\varepsilon}$, ${p}$, ${u}$, $\dot{u}$). 
The representation, Eq. \ref{EBloch}, 
separates the time harmonic and macroscopic factor from the microscopic part of the field variables. 
Even for a finite unit cell, the Fourier series solution of the microscopic part is periodic so that this solution satisfies,
$F(x)=F(x+m^{'}a)$. 
We emphasize here that the frequency and the wavenumber, $\omega$ and $q$, are, at this point, unrelated and arbitrary.  

The local conservation and kinematic relations are,

\begin{equation}
\begin{array}{l}
\displaystyle \nabla\sigma=-i\omega p\\
\displaystyle \nabla \dot{u}=-i\omega\varepsilon
\end{array} 
\label{EFieldEqn}
\end{equation}

where $\nabla\rightarrow\frac{\partial}{\partial x}+iq $.  
The corresponding constitutive relations are,

\begin{equation}
\begin{array}{l}
\displaystyle \varepsilon=D(x)\sigma\\
\displaystyle p=\rho(x) \dot{u}
\end{array} 
\label{EFieldEqn1}
\end{equation}

where $D(x)$ is the compliance, the inverse of the elastic modulus, and $\rho(x)$ is the mass-density.  These local material parameters represent the structure and composition of the unit cell.

Now we replace the heterogeneous unit cell with a homogeneous one having uniform density $\rho_0$ and compliance $D_0$. 
In order to reproduce the strain and momentum of the actual unit cell, field variables eigenstress, 
${\tilde{\Sigma} (x)}$, and eigenvelocity, 
${\tilde{\dot{U}}(x)}$, are introduced.  These quantities are then calculated, using the basic local field equations and constitutive relations.
The idea stems from the polarization stress or strain that was originally proposed by Hashin \cite{Hashin1} and further developed by Hashin and Shtriktman \cite{Hashin2,  Hashin3} and later by others, in order to construct energy-based bounds for the composite's overall elastic moduli. 
The basic tool in these works has been the result obtained by Eshelby \cite{Eshelby} in three dimensions and earlier by Hardiman \cite{Hardiman} in two dimensions, 
that the stress and strain are constant within an ellipsoidal (elliptical in two dimensions) region of an infinitely extended uniform elastic medium when that region undergoes a uniform transformation corresponding to a uniform inelastic strain. 

Here, however, we present a different tool that can be used to actually calculate the point-wise values of the elasto-dynamic field variables to any desired degree of accuracy.  
For this, we require that the actual values of the field variables at every point within the homogenized and the original heterogeneous unit cell be exactly the same.  
To ensure this, we require that the following consistency conditions hold at every point within the unit cell:

\begin{equation}
\begin{array}{l}
\displaystyle \varepsilon=D\sigma=D_0(\sigma-\tilde{\Sigma})\\
\displaystyle p=\rho \dot{u}=\rho_0(\dot{u}-\tilde{\dot{U}})
\end{array} 
\label{Eeigenstuff}
\end{equation}

The eigenstress and eigenvelocity fields are zero in regions where the material properties of the heterogeneous unit cell are equal to the chosen uniform material properties, 
$D_0$ and $\rho_0$. 
From Eqs. (\ref{EFieldEqn}, \ref{Eeigenstuff}) we have,

\begin{equation}\label{Ee1}
\nabla^2\sigma+\nu^2\sigma=\nu^2\tilde{\Sigma}-\frac{\nu^2}{i\omega D_0}\nabla\tilde{\dot{U}}
\end{equation}

\begin{equation}\label{Ee2}
\nabla^2\dot{u}+\nu^2\dot{u}=\nu^2\tilde{\dot{U}}-\frac{\nu^2}{i\omega \rho_0}\nabla\tilde{\Sigma}
\end{equation}

where $\nu^2=\omega^2\rho_0D_0$. 
We now consider a Fourier series solution of the above equations. 
With $\xi=\pm 2n\pi/a,n\neq 0$, we set,

\begin{equation}%\label{EBloch}
F(x)=F^0+F^p=\langle F\rangle_\Omega+\sum_{\xi\neq 0}F(\xi)\mathrm{e}^{i\xi x}
\end{equation}

\begin{equation}\label{EF}
\langle F\rangle_\Omega=\frac{1}{\Omega}\int_\Omega F(x)dx
\end{equation}

\begin{equation}%\label{EBloch}
F(\xi)=\frac{1}{\Omega}\int_\Omega F(x)\mathrm{e}^{-i\xi x} dx
\end{equation}

where $\langle F\rangle_\Omega$ represents the averaged value of the field variable, {F}, over the unit cell, and  $F(\xi)$ represents the Fourier coefficient of the corresponding local perturbation due to heterogeneity. 
From Eqs. (\ref{Ee1},  \ref{Ee2}) we now obtain the following Fourier coefficients for the stress and velocity fields:

\begin{equation}
\sigma(\xi)=\frac{\nu^2}{[\nu^2-(\xi+q)^2]}\tilde{{\Sigma}}(\xi)-\frac{\nu^2(\xi+q)}{\omega D_0[\nu^2-(\xi+q)^2]}\tilde{\dot{U}}(\xi)
\end{equation}

\begin{equation}
\dot{u}(\xi)=\frac{\nu^2}{[\nu^2-(\xi+q)^2]}\tilde{\dot{U}}(\xi)-\frac{\nu^2(\xi+q)}{\omega\rho_0[\nu^2-(\xi+q)^2]}\tilde{\Sigma}(\xi)
\end{equation}

Therefore, the stress and velocity fields can be expressed as a sum of their average and their periodic components,

\begin{equation}\label{EStressPer}
\sigma(x)=\langle \sigma\rangle+\sum_{\xi\neq 0}e^{i\xi x}\left[A(\xi)\frac{1}{\Omega}\int_{\Omega}\tilde{\Sigma}(y)e^{-i\xi y}dy-\frac{B(\xi)}{\omega D_0}\frac{1}{\Omega}\int_{\Omega}\tilde{\dot{U}}(y)e^{-i\xi y}dy\right]
\end{equation}

\begin{equation}\label{EDispPer}
\dot{u}(x)=\langle \dot{u}\rangle+\sum_{\xi\neq 0}e^{i\xi x}\left[A(\xi)\frac{1}{\Omega}\int_{\Omega}\tilde{\dot{U}}(y)e^{-i\xi y}dy-\frac{B(\xi)}{\omega\rho_0}\frac{1}{\Omega}\int_{\Omega}\tilde{\Sigma}(y)e^{-i\xi y}dy\right]
\end{equation}

\begin{equation}\label{A_B}
A(\xi)=\frac{\nu^2}{[\nu^2-(\xi+q)^2]}; \quad B(\xi)=\frac{\nu^2(\xi+q)}{[\nu^2-(\xi+q)^2]}
\end{equation}

where $\langle \dot{u}\rangle$ and $\langle \sigma\rangle$ are the average values of the velocity and stress fields, respectively, taken over a unit cell. Note that $B(\xi)$ has the form,

\begin{equation}\label{B}
B(\xi)={\xi}A(\xi)+qA(\xi)
\end{equation}

To make the homogenized unit cell point-wise equivalent to the original heterogeneous unit cell, the homogenizing fields are required to satisfy the following consistency conditions:

\begin{equation}
D(x)[\langle\sigma\rangle+\sigma^p]=D_0[\langle\sigma\rangle+\sigma^p-\tilde{\Sigma}]
\end{equation}

\begin{equation}
\rho(x)[\langle \dot{u}\rangle+\dot{u}^p]=\rho_0[\langle \dot{u}\rangle+\dot{u}^p-\tilde{\dot{U}}]
\end{equation}

The periodic parts of the velocity and stress fields, from Eqs. (\ref{EDispPer}, \ref{EStressPer}), are now substituted into the above equations. 
This gives a set of 2 coupled integral equations which yields the required homogenizing stress and velocity fields that exactly and fully replace the heterogeneity in the original medium. 
Our immediate concern, however, is not the point-wise representation of the heterogeneous medium, but, rather, it is the determination of the averaged field values, though the solution technique also yields the point-wise values of the field values as well. 

 We now divide the unit cell into $\bar{\alpha}$ number of subregions, $\Omega_\alpha$. 
 Then, we average the perturbation fields over each such subregion to obtain,

\begin{eqnarray}\label{EStressXiAv}
\lefteqn{\langle \sigma^p\rangle_{\Omega_\alpha}=\sigma^p_\alpha=\frac{1}{\Omega_\alpha}\int_{\Omega_\alpha}\sigma^p(x)dx}\\
& & \nonumber = \sum_{\xi\neq 0}g_\alpha(\xi)\left( A(\xi) \frac{1}{\Omega}\int_\Omega
\tilde{\Sigma}(y)\mathrm{e}^{-i\xi y}dy - \frac{B(\xi)}{\omega D_0}\frac{1}{\Omega}\int_\Omega
\tilde{\dot{U}}(y)\mathrm{e}^{-i\xi y}dy \right)
\end{eqnarray}

\begin{eqnarray}\label{EDisplacementXiAv}
\lefteqn{\langle \dot{u}^p\rangle_{\Omega_\alpha}=\dot{u}^p_\alpha=\frac{1}{\Omega_\alpha}\int_{\Omega_\alpha}\dot{u}^p(x)dx}\\
& & \nonumber = \sum_{\xi\neq 0}g_\alpha(\xi)\left( A(\xi) \frac{1}{\Omega}\int_\Omega
\tilde{\dot{U}}(y)\mathrm{e}^{-i\xi y}dy - \frac{B(\xi)}{\omega\rho_0}\frac{1}{\Omega}\int_\Omega
\tilde{\Sigma}(y)\mathrm{e}^{-i\xi y}dy \right)
\end{eqnarray}

\begin{equation}%\label{EStressXi}
g_\alpha(\xi)=\frac{1}{\Omega_\alpha}\int_{\Omega_\alpha}\mathrm{e}^{i\xi x}dx
\end{equation}

We now replace the integrals in Eqs. (\ref{EStressXiAv}, \ref{EDisplacementXiAv}) by their equivalent finite sums and set,

\begin{equation}
\begin{array}{l}
\displaystyle \frac{1}{\Omega}\int_\Omega F(y)\mathrm{e}^{-i\xi y}dy\approx \sum_\beta f_\beta g_\beta(-\xi)F_\beta\\
\displaystyle f_\beta=\frac{\Omega_\beta}{\Omega}\\
\displaystyle F_\beta=\langle F\rangle_{\Omega_\beta}
\end{array} 
%\label{EFieldEqn}
\end{equation}

Eqs. (\ref{EStressXiAv}, \ref{EDisplacementXiAv}) then yield the following expressions:

\begin{equation}\label{EStressXiAvbeta}
\sigma^p_\alpha=\tilde{A}_{\alpha\beta}\tilde{\Sigma}_\beta-\frac{1}{\omega D_0}\tilde{B}_{\alpha\beta}\tilde{\dot{U}}_\beta
\end{equation}

\begin{equation}\label{EDisplacementXiAvbeta}
\dot{u}^p_\alpha=\tilde{A}_{\alpha\beta}\tilde{\dot{U}}_\beta-\frac{1}{\omega\rho_0}\tilde{B}_{\alpha\beta}\tilde{\Sigma}_\beta
\end{equation}

where the repeated index, $\beta$, is summed over the number of subregions, ${\beta} = 1, \dots, \bar{\alpha}$.  The coefficient matrices in the above equations are defined by,

\begin{equation}
\begin{array}{l}
\displaystyle \tilde{A}_{\alpha\beta}=\sum_{\xi\neq 0}g_\alpha(\xi)f_\beta g_\beta(-\xi)A(\xi)\\
\displaystyle \tilde{B}_{\alpha\beta}=\sum_{\xi\neq 0}g_\alpha(\xi)f_\beta g_\beta(-\xi)B(\xi)
\end{array} 
%\label{EFieldEqn}
\end{equation}

In these equations, $\beta$ is not summed. Averaging the consistency conditions over each subregion $\alpha$ and using Eqs. (\ref{EStressXiAvbeta}, \ref{EDisplacementXiAvbeta}), we have,

\begin{equation}\label{EFinal1}
f_\alpha\langle\sigma\rangle=-\left[\bar{A}_{\alpha\beta}+\frac{f_\alpha D_0}{D_\alpha-D_0}\delta_{\alpha\beta}\right]\tilde{\Sigma}_\beta+\frac{1}{\omega D_0}\bar{B}_{\alpha\beta}\tilde{\dot{U}}_\beta 
\end{equation}

\begin{equation}\label{EFinal2}
f_\alpha\langle \dot{u}\rangle =\frac{1}{\omega\rho_0}\bar{B}_{\alpha\beta}\tilde{\Sigma}_\beta- \left[\bar{A}_{\alpha\beta}+\frac{f_\alpha\rho_0}{\rho_\alpha-\rho_0}\delta_{\alpha\beta}\right]\tilde{\dot{U}}_\beta
\end{equation}

\begin{equation}
\begin{array}{l}
\displaystyle \bar{A}_{\alpha\beta}=f_\alpha\tilde{A}_{\alpha\beta}\\
\displaystyle \bar{B}_{\alpha\beta}=f_\alpha\tilde{B}_{\alpha\beta}\quad\mathrm{\alpha \;not \;summed}\\
\end{array} 
\end{equation}

Note that $\bar{A}_{\alpha\beta}$ and $\bar{B}_{\alpha\beta}$ are geometric quantities,  independent of the material properties of the unit cell.  Expressions (\ref{EFinal1}, \ref{EFinal2}) are $2\bar{\alpha}$ linear equations which can be solved for $\bar{\alpha}$ number of $\tilde{\Sigma}_\beta$ and  $\bar{\alpha}$ number of $\tilde{\dot{U}}_\beta$ in terms of the average stess $\langle\sigma\rangle$ and average velocity $\langle \dot{u} \rangle$.  
We express the solution in the following matrix form:

\begin{equation}
\begin{array}{l}
\displaystyle \{\tilde{\boldsymbol{\Sigma}}\}=\{\boldsymbol{\Phi}\}\langle\sigma\rangle+\frac{1}{D_0}\{\boldsymbol{\Psi}\}\langle \dot{u}\rangle\\
\displaystyle \{\tilde{\dot{\mathbf{U}}}\}=\frac{1}{\rho_0}\{\boldsymbol{\Theta}\}\langle\sigma\rangle+\{\boldsymbol{\Gamma}\}\langle \dot{u}\rangle\\
\end{array} 
%\label{EFieldEqn}
\end{equation}

where,

\begin{equation}
\begin{array}{l}
\displaystyle \{\boldsymbol{\Phi}\}=\left[-\left[\mathbf{A_D}\right]+\frac{1}{\nu^2}\left[\mathbf{B}\right]\left[\mathbf{A_\rho}\right]^{-1}\left[\mathbf{B}\right]\right]^{-1}\{f\}\\
\displaystyle \{\boldsymbol{\Psi}\}=\frac{1}{\omega}\left[-\left[\mathbf{A_D}\right]+\frac{1}{\nu^2}\left[\mathbf{B}\right]\left[\mathbf{A_\rho}\right]^{-1}\left[\mathbf{B}\right]\right]^{-1}\left[\mathbf{B}\right]\left[\mathbf{A_\rho}\right]^{-1}\{f\}\\

\displaystyle \{\boldsymbol{\Theta}\}=\frac{1}{\omega}\left[-\left[\mathbf{A_\rho}\right]+\frac{1}{\nu^2}\left[\mathbf{B}\right]\left[\mathbf{A_D}\right]^{-1}\left[\mathbf{B}\right]\right]^{-1}\left[\mathbf{B}\right]\left[\mathbf{A_D}\right]^{-1}\{f\}\\
\displaystyle \{\boldsymbol{\Gamma}\}=\left[-\left[\mathbf{A_\rho}\right]+\frac{1}{\nu^2}\left[\mathbf{B}\right]\left[\mathbf{A_D}\right]^{-1}\left[\mathbf{B}\right]\right]^{-1}\{f\}\\

\displaystyle \left[\mathbf{A_D}\right]_{\alpha\beta}=\bar{A}_{\alpha\beta}+\frac{f_\alpha D_0}{D_\alpha-D_0}\delta_{\alpha\beta}\\

\displaystyle \left[\mathbf{A_\rho}\right]_{\alpha\beta}=\bar{A}_{\alpha\beta}+\frac{f_\alpha\rho_0}{\rho_\alpha-\rho_0}\delta_{\alpha\beta}\\
\displaystyle \{f\}^T=\{f_1,f_2,...f_{\bar{\alpha}}\}\\
\end{array} 
\label{EFinalConst}
\end{equation}

Now we average the consistency conditions over the entire unit cell to express the average strain and average momentum in terms of the average stress and average velocity. Noting that the average of the periodic parts vanish when taken over the entire unit cell, we have,

\begin{equation}\label{e15}
\begin{array}{l}
\displaystyle \langle\varepsilon\rangle=D_0[\langle\sigma\rangle-\langle\tilde{\Sigma}\rangle]\\
\displaystyle \quad\;\;=\bar{D}\langle\sigma\rangle+\bar{S}_1\langle \dot{u}\rangle\\
\end{array} 
\end{equation}

\begin{equation}\label{e16}
\begin{array}{l}
\displaystyle \langle p\rangle=\rho_0[\langle \dot{u}\rangle-\langle\tilde{\dot{U}}\rangle]\\
\displaystyle \quad\;\;=\bar{S}_2\langle\sigma\rangle+\bar{\rho}\langle \dot{u}\rangle\\
\end{array} 
\end{equation}

where

\begin{equation}
\begin{array}{l}
\displaystyle \bar{D}=D_0[1-\{f\}^T\{\boldsymbol{\Phi}\}]\\
\displaystyle \bar{S}_1=-D_0\{f\}^T\frac{1}{D_0}\{\boldsymbol{\Psi}\}=-\{f\}^T\{\boldsymbol{\Psi}\}\\
\displaystyle \bar{S}_2=-\rho_0\{f\}^T\frac{1}{\rho_0}\{\boldsymbol{\Theta}\}=-\{f\}^T\{\boldsymbol{\Theta}\}\\
\displaystyle \bar{\rho}=\rho_0[1-\{f\}^T\{\boldsymbol{\Gamma}\}]\\
\end{array} 
\label{EEffective}
\end{equation}

Eqs. (\ref{e15}, \ref{e16}) are our final constitutive relations for the homogenized composite. In these equations, $\bar{D}$ and $\bar{\rho}$ are the overall homogenized composite's compliance and mass-density, respectively.  \emph{As is shown below, $\bar{D}$ and $\bar{\rho}$ are alway real-valued and positive.  
The coupling terms, $\bar{S_1}$ and $\bar{S_2}$, are always each others' complex conjugate.  These attributes hold whether the unit cell is symmetric or non-symmetric.} The coupling terms of course vanish if the unit cell is homogeneous. They are real-valued and equal when the unit cell has reflective symmetry.

\subsection{Mathematical Structure of Constitutive Relations}

We note  that the matrices $\left[\mathbf{A_D}\right]$, $\left[\mathbf{A_\rho}\right]$, and $\left[\mathbf{B}\right]$, given by,

\begin{equation}
\begin{array}{l}
\displaystyle \left[\mathbf{A_D}\right]_{\alpha\beta}=\sum_{\xi>0}\left[f_\alpha g_\alpha(\xi)f_\beta g_\beta(-\xi)A(\xi)+f_\alpha g_\alpha(-\xi)f_\beta g_\beta(\xi)A(-\xi)\right]+\frac{f_\alpha D_0}{D_\alpha-D_0}\delta_{\alpha\beta}\\

\displaystyle \left[\mathbf{A_\rho}\right]_{\alpha\beta}=\sum_{\xi>0}\left[f_\alpha g_\alpha(\xi)f_\beta g_\beta(-\xi)A(\xi)+f_\alpha g_\alpha(-\xi)f_\beta g_\beta(\xi)A(-\xi)\right]+\frac{f_\alpha \rho_0}{\rho_\alpha-\rho_0}\delta_{\alpha\beta}\\

\displaystyle \left[\mathbf{B}\right]_{\alpha\beta}=\sum_{\xi>0}\left[f_\alpha g_\alpha(\xi)f_\beta g_\beta(-\xi)B(\xi)+f_\alpha g_\alpha(-\xi)f_\beta g_\beta(\xi)B(-\xi)\right]\\

\end{array} 
%\label{EFieldEqn}
\end{equation}

are Hermitian for real values of the wavenumber. For any Hermitian matrix $\left[\mathbf{M}\right]$, one has,

\begin{equation}\nonumber
\left[\left[\mathbf{M}\right]^{-1}\right]^*=\left[\left[\mathbf{M}\right]^*\right]^{-1}=\left[\mathbf{M}\right]^{-1}
\end{equation}

where asterix denotes a Hermitian transpose. Using this and the basic properties of transpose of products of matrices, it follows that both $\left[-\left[\mathbf{A_D}\right]+\frac{1}{\nu^2}\left[\mathbf{B}\right]\left[\mathbf{A_\rho}\right]^{-1}\left[\mathbf{B}\right]\right]^{-1}$ and $\left[-\left[\mathbf{A_\rho}\right]+\frac{1}{\nu^2}\left[\mathbf{B}\right]\left[\mathbf{A_D}\right]^{-1}\left[\mathbf{B}\right]\right]^{-1}$ are Hermitian matrices. Moreover, for a Hermitian matrix $\left[\mathbf{M}\right]$, it can be shown that the scalar $\{f\}^T\left[\mathbf{M}\right]\{f\}$ is always real-valued. Therefore, from Eq. (\ref{EFinalConst}), 

\begin{equation}
\begin{array}{l}
\displaystyle \bar{D}=D_0\left[1-\{f\}^T\left[-\left[\mathbf{A_D}\right]+\frac{1}{\nu^2}\left[\mathbf{B}\right]\left[\mathbf{A_\rho}\right]^{-1}\left[\mathbf{B}\right]\right]^{-1}\{f\}\right]\\
\displaystyle \bar{\rho}=\rho_0\left[1-\{f\}^T\left[-\left[\mathbf{A_\rho}\right]+\frac{1}{\nu^2}\left[\mathbf{B}\right]\left[\mathbf{A_D}\right]^{-1}\left[\mathbf{B}\right]\right]^{-1}\{f\}\right]\\

\end{array} 
%\label{EFieldEqn}
\end{equation}

are always real-valued. We also have the following identities:

\begin{eqnarray}
  \bar{S}_1^* & = & -\left[\{f\}^T\{\boldsymbol{\Psi}\}\right]^*\nonumber \\
   & = & -\frac{1}{\omega}\left[\{f\}^T\left[-\left[\mathbf{A_D}\right]+\frac{1}{\nu^2}\left[\mathbf{B}\right]\left[\mathbf{A_\rho}\right]^{-1}\left[\mathbf{B}\right]\right]^{-1}\left[\mathbf{B}\right]\left[\mathbf{A_\rho}\right]^{-1}\{f\}\right]^* \nonumber \\
   & = & -\frac{1}{\omega}\left[\left[\left[\mathbf{B}\right]\left[\mathbf{A_\rho}\right]^{-1}\{f\}\right]^*\left[\{f\}^T\left[-\left[\mathbf{A_D}\right]+\frac{1}{\nu^2}\left[\mathbf{B}\right]\left[\mathbf{A_\rho}\right]^{-1}\left[\mathbf{B}\right]\right]^{-1}\right]^*\right] \nonumber \\
   & = & -\frac{1}{\omega}\left[\{f\}^T\left[\mathbf{A_\rho}\right]^{-1}\left[\mathbf{B}\right]\left[-\left[\mathbf{A_D}\right]+\frac{1}{\nu^2}\left[\mathbf{B}\right]\left[\mathbf{A_\rho}\right]^{-1}\left[\mathbf{B}\right]\right]^{-1}\{f\}\right] \nonumber \\
   & = & -\frac{1}{\omega}\left[\{f\}^T\left[\left[-\left[\mathbf{A_D}\right]+\frac{1}{\nu^2}\left[\mathbf{B}\right]\left[\mathbf{A_\rho}\right]^{-1}\left[\mathbf{B}\right]\right]\left[\mathbf{B}\right]^{-1}\left[\mathbf{A_\rho}\right]\right]^{-1}\{f\}\right] \nonumber \\
   & = & -\frac{1}{\omega}\left[\{f\}^T\left[-\left[\mathbf{A_D}\right]\left[\mathbf{B}\right]^{-1}\left[\mathbf{A_\rho}\right]+\frac{1}{\nu^2}\left[\mathbf{B}\right]\right]^{-1}\{f\}\right] \nonumber \\
\end{eqnarray}

and,

\begin{eqnarray}
  \bar{S}_2 & = & -\left[\{f\}^T\{\boldsymbol{\Theta}\}\right]\nonumber \\
   & = & -\frac{1}{\omega}\left[\{f\}^T\left[-\left[\mathbf{A_\rho}\right]+\frac{1}{\nu^2}\left[\mathbf{B}\right]\left[\mathbf{A_D}\right]^{-1}\left[\mathbf{B}\right]\right]^{-1}\left[\mathbf{B}\right]\left[\mathbf{A_D}\right]^{-1}\{f\}\right] \nonumber \\
   & = & -\frac{1}{\omega}\left[\{f\}^T\left[\left[\mathbf{A_D}\right]\left[\mathbf{B}\right]^{-1}\left[-\left[\mathbf{A_\rho}\right]+\frac{1}{\nu^2}\left[\mathbf{B}\right]\left[\mathbf{A_D}\right]^{-1}\left[\mathbf{B}\right]\right]\right]^{-1}\{f\}\right] \nonumber \\
   & = & -\frac{1}{\omega}\left[\{f\}^T\left[-\left[\mathbf{A_D}\right]\left[\mathbf{B}\right]^{-1}\left[\mathbf{A_\rho}\right]+\frac{1}{\nu^2}\left[\mathbf{B}\right]\right]^{-1}\{f\}\right] \end{eqnarray}

which show that $\bar{S}_1^*=\bar{S}_2$. It can be seen that the matrices $\left[\mathbf{A_D}\right]$, $\left[\mathbf{A_\rho}\right]$, and $\left[\mathbf{B}\right]$ are Hermitian regardless of the  asymmetric nature of the unit cell. While asymmetries related to dimensions affect the off diagonal terms in a way which preserves the Hermitian nature of the matrices, asymmetries related to material properties only affect the diagonal terms, thereby, preserving the Hermitian characteristics. This in effect means that, for any asymmetrical unit cell, $\bar{D}$ and $\bar{\rho}$ are always real-valued and that the coupling terms in (\ref{e15}, \ref{e16}) are always complex conjugates of each other. This also ensures that the total energy given by,

\begin{eqnarray}
  E & = &  \frac{1}{4}\left[\langle\sigma\rangle^*\langle\varepsilon\rangle+\langle\sigma\rangle\langle\varepsilon\rangle^*+\langle\dot{u}\rangle^*\langle p\rangle+\langle\dot{u}\rangle\langle p\rangle^*\right]\\
   & = & \frac{1}{4}\left[2\bar{D}\langle\sigma\rangle\langle\sigma\rangle^*+2\bar{\rho}\langle\dot{u}\rangle\langle\dot{u}\rangle^*+2\bar{S}_2\langle\sigma\rangle\langle\dot{u}\rangle^*+2\bar{S}_1\langle\sigma\rangle^*\langle\dot{u}\rangle\right]\nonumber \\
   & = & \frac{1}{4}\left[2\bar{D}\langle\sigma\rangle\langle\sigma\rangle^*+2\bar{\rho}\langle\dot{u}\rangle\langle\dot{u}\rangle^*+2\bar{S}_1^*\langle\sigma\rangle\langle\dot{u}\rangle^*+2\bar{S}_1\langle\sigma\rangle^*\langle\dot{u}\rangle\right]\nonumber \nonumber \\
   & = & \frac{1}{2}\left[\bar{D}\langle\sigma\rangle\langle\sigma\rangle^*+\bar{\rho}\langle\dot{u}\rangle\langle\dot{u}\rangle^*+\left[\bar{S}_1\langle\sigma\rangle^*\langle\dot{u}\rangle\right]^*+\bar{S}_1\langle\sigma\rangle^*\langle\dot{u}\rangle\right]\nonumber \end{eqnarray}

is always real.  Now, since $E$ is also non-negative, and that the average stress and velocity can be prescribed essentially arbitrarily and independently, it follows that $\bar{D}$ and $\bar{\rho}$ are also positive.  

\subsection{Bloch-form Overall Spatial Variation and Dispersion Relations}  

We now consider a special case of an infinite homogenized elastic solid with a layered microstructure, and seek conditions under which it supports periodic waves of the Bloch-form spatial variation of the following form:

\begin{equation}
\begin{array}{l}\label{B_Sp}
\displaystyle \langle\sigma\rangle(x)=\langle\sigma\rangle e^{iqx};\quad \langle \dot{u}\rangle(x)=\langle \dot{u}\rangle e^{iqx}\\
\displaystyle \langle\varepsilon\rangle(x)=\langle\varepsilon\rangle e^{iqx};\quad \langle p\rangle(x)=\langle p\rangle e^{iqx}\\
\end{array} 
%\label{EFieldEqn}
\end{equation}

The overall field equations then become,

\begin{equation}
\begin{array}{l}
\displaystyle \frac{d}{dx}(\langle\sigma\rangle e^{iqx})=-i\omega\langle p\rangle e^{iqx};\quad
\displaystyle  \frac{d}{dx}(\langle \dot{u}\rangle e^{iqx})=-i\omega\langle \varepsilon\rangle e^{iqx}\\
\end{array} 
%\label{EFieldEqn}
\end{equation}

which yield,

\begin{equation}\label{B_Field}
q\langle\sigma\rangle=-\omega\langle p\rangle;\quad q\langle \dot{u}\rangle=-\omega\langle \varepsilon\rangle
\end{equation}

From these equations and the overall constitutive relations (\ref{e15}, \ref{e16}) it now follows that, for this homogenized continuum to admit solutions of the form (\ref{B_Sp}),  the following dispersion relation must necessarily be satisfied:

\begin{equation}\label{Dis_Ov}
\left(\frac{\omega}{q}\right)^2=v_p^2=\frac{(1+v_p\bar{S}_1)(1+v_p\bar{S}_2)}{\bar{D}\bar{\rho}}
\end{equation}

where $v_p$ is the phase velocity. The above equation is used to evaluate the dispersion relation for a layered composite by substituting from Eq. (\ref{EEffective}). The real frequency-wavenumber pairs are then used to evaluate the effective parameters of Eq. (\ref{e15}, \ref{e16}) and these parameters would have the properties discussed in the preceding subsection. 

If we wish, as in \cite{NWSA}, to directly relate the overall strain, $\langle \varepsilon\rangle$, to the overall stress, 
$\langle\sigma\rangle$, and the overall momentum,
$\langle{p}\rangle$, to the overall velocity, $\langle\dot{u}\rangle$,  according to

\begin{equation}
\begin{array}{l}
\displaystyle \langle\varepsilon\rangle=D^\mathrm{eff}\langle\sigma\rangle\\
\displaystyle \langle p\rangle=\rho^\mathrm{eff}\langle\dot{u}\rangle\\
\end{array} 
\label{EEffective2}
\end{equation}

then we have,

\begin{equation}\label{EDeff}
D^\mathrm{eff}=\frac{\bar{D}}{1+v_p\bar{S}_1}
\end{equation}

\begin{equation}\label{Erhoeff}
\rho^\mathrm{eff}=\frac{\bar{\rho}}{1+v_p\bar{S}_2}
\end{equation}

and they do satisfy the same dispersion relation Eq. (\ref{Dis_Ov}), now rewritten as,

\begin{equation}\label{Dis_Ov1}
\left(\frac{\omega}{q}\right)^2=v_p^2=\frac{1}{D^\mathrm{eff}\rho^\mathrm{eff}}
\end{equation}

In this representation however, the effective parameters defined by (\ref{EEffective2}) are real only for symmetric (elastic) unit cells and become complex-valued for asymmetric unit cells. Also the parameters $D^\mathrm{eff}$ and $\rho^\mathrm{eff}$ are obtained using the special form of the field equations (\ref{B_Field}) while the constitutive relations (\ref{e15}, \ref{e16}) have broader applicability.

A significant advantage of our homogenization technique is that the point-wise solution of the field equations is not required \emph{a priori}, and that, this solution can be obtained as a by product of the approach.
In fact, within each subregion, the value of each field variable is given by adding to each periodic part in Eqs. (\ref{EStressXiAvbeta}, \ref{EDisplacementXiAvbeta}), the corresponding average value.  
We do not need explicit expressions for the mode-shapes, but rather can extract these from the final results, if desired.  
The dispersion relations are obtained by substituting Eqs. (\ref{e15}, \ref{e16}) into Eq. (\ref{Dis_Ov}).  These relations then relate the frequency, $\omega$, and the wavenumber, $q$, for the special Bloch-form of the periodic waves, as exemplified in what follows.

\subsection{Example: Dispersion calculation for a layered composite}

\begin{figure}[htp]
\centering
\includegraphics[scale=.8]{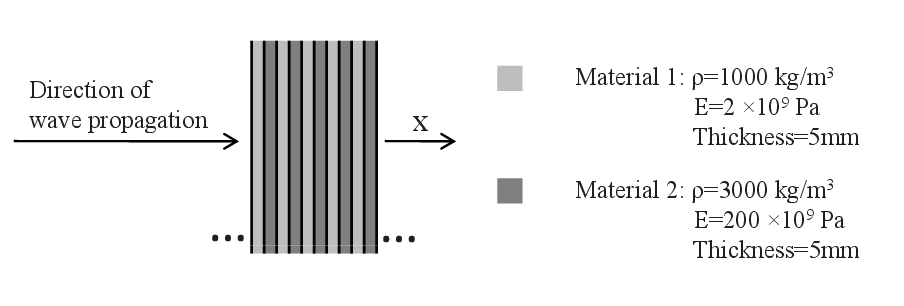}
\caption{Schematic of a bi-layered composite}\label{FLayered}
\end{figure}

To illustrate dispersion calculations by our micro-mechanical formulation, consider the case of a layered composite (Fig. \ref{FLayered}) with harmonic waves traveling perpendicular to the layers.

The exact dispersion relation for longitudinal waves in a layered composite was given by Rytov \cite{Rytov}. Here we  compare the dispersion curves calculated using our micro-mechanical formulation with those of the exact solution. A unit cell, for this case, consists of 1 layer of Material 1 and 1 layer of Material 2. We divide the unit cell into $N=30$ subregions (15 divisions per layer) and use constant values for eigenstress and eigenvelocity within each subregion. 

\begin{figure}[htp]
\centering
\includegraphics[scale=.6]{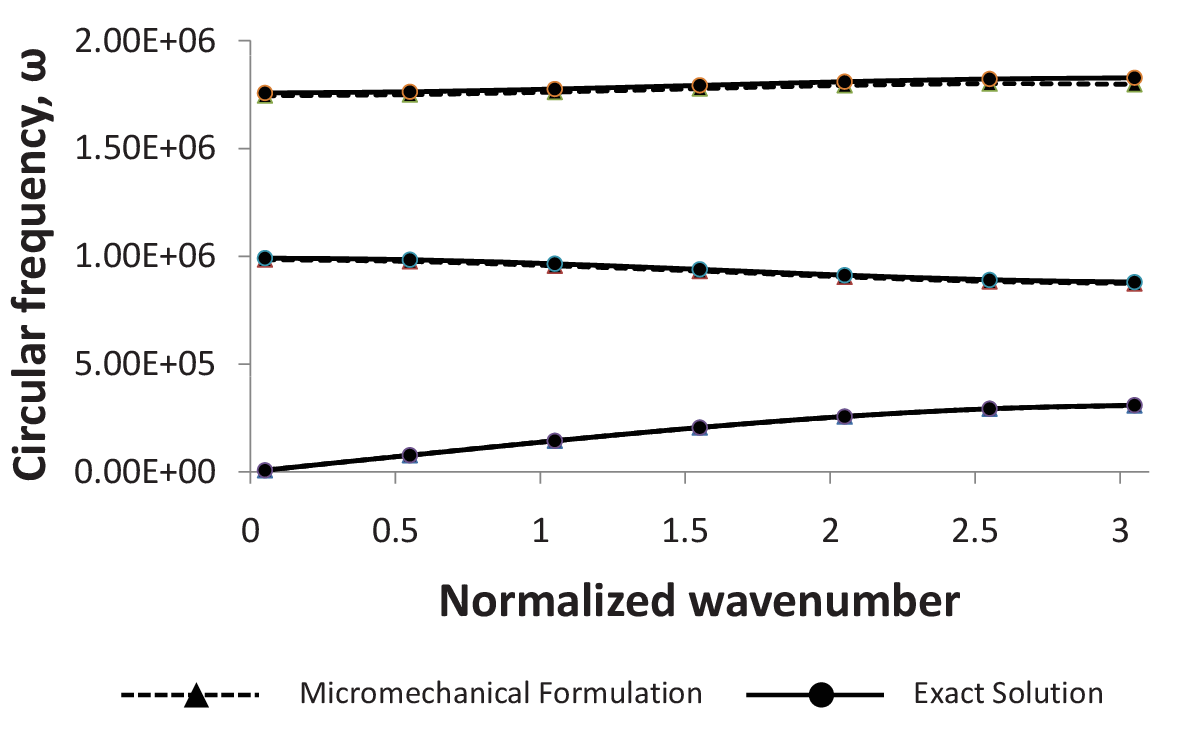}
\caption{Dispersion curve comparison of the Micro-mechanical formulation with the exact results}\label{FDispersionComp}
\end{figure}

The first three modes in the dispersion curve are compared in Fig. (\ref{FDispersionComp}). It can be seen that the micro-mechanical formulation gives very acceptable results for the first three modes when appropriate discretization is used. There are indications that the results of the micro-mechanical formulation start diverging from the exact results as one considers higher frequencies. This is evidenced by the increasing difference in the results as one moves up the third branch. This is of course an issue of discretization and the micro-mechanical formulation is expected to give accurate dispersion results at any frequency given a high enough discretization. 

It should be mentioned here that exact dispersion relations are only available for fairly simple geometries like layered composites. The lack of exact solutions for more complex cases, therefore, necessitates accurate and efficient numerical schemes. In a recent paper \cite{NWSA}, a mixed method is used to obtain the point-wise values of the field variables and  to formulate the corresponding integration-based homogenization of layered composites. This method was shown to be in good agreement with homogenization based upon the integration of field variables as calculated from the exact solution. Although homogenization based on the mixed variational method  provides an accurate and efficient numerical alternative to using the exact solution, the field variables still need to be evaluated at multiple points within the unit cell. Here we show that the micro-mechanical approach to homogenization, in addition to precluding the need for point-wise evaluation of the field variables, also converges to homogenization results based on both the mixed variational method and the exact solution; a brief summary of integration of field variable based homogenization is presented in Appendix \ref{Appendix}. 

\section{Constitutive relations of layered composites}

To illustrate the micro-structural method of calculating the constitutive parameters of homogenized composites, examples of  4-layered symmetric and asymmetric composites are considered. These examples also serve to show that homogenization based on the micro-structural formulation converges to the results of homogenization based on the integration of field variables using the exact solution and/or using the results of the mixed variational method.

\subsection{4-layered symmetric composite: Comparison with homogenization based on the mixed variational formulation}

\begin{figure}[htp]
\centering
\includegraphics[scale=.75]{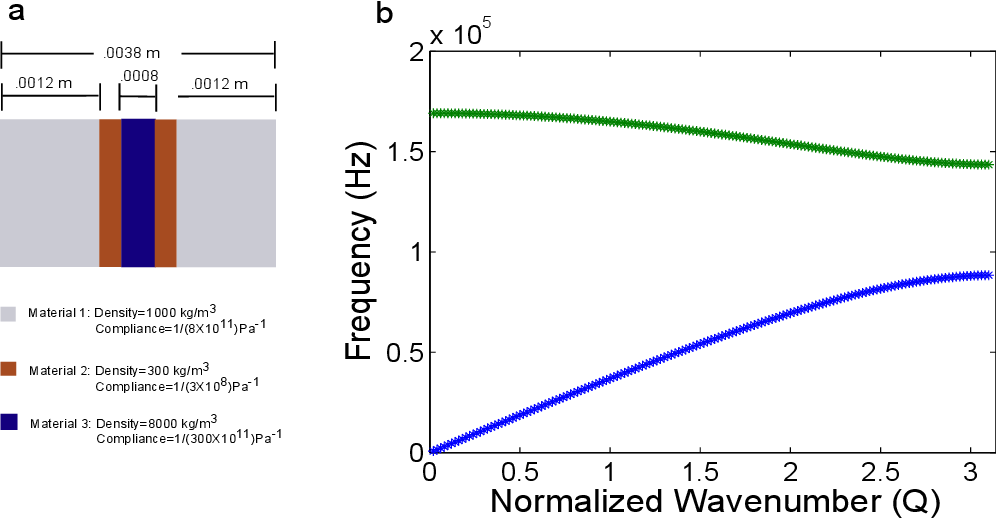}
\caption{4-layered symmetric composite: a. Schematic of a unit cell; b. Dispersion curve for the first 2 propagating branches}\label{FUnitCell4Layered}
\end{figure}

We consider a 4-layered composite in order to show that the micro-structural homogenization gives results which converge to homogenization based on the integration of field variables as derived from the mixed variational formulation. We stressed that for this integration-based homogenization, the effective parameters are calculated from Eqs. (\ref{EEffectiveC}, \ref{EEffectiveRho}) using the mode-shapes calculated from the approximate mixed variational formulation of \cite{SNN1}, whereas for the micro-structural homogenization the constitutive parameters are calculated from Eqs. (\ref{EDeff}, \ref{Erhoeff}) directly. It should be noted that while the effective parameters calculated from Eqs. (\ref{EEffectiveC}, \ref{EEffectiveRho}) require integration of the field variables through the unit cell, effective parameters calculated from Eqs. (\ref{EDeff}, \ref{Erhoeff}) do not require any integration. For symmetric unit cells such as Fig. (\ref{FUnitCell4Layered}a), Eqs. (\ref{EDeff}, \ref{Erhoeff}) result in real valued effective parameters and produce the same constitutive relations (\ref{EEffectiveC}, \ref{EEffectiveRho}) reported in literature \cite{NWSA}. Expressions (\ref{EEffectiveC}, \ref{EEffectiveRho}) however do not clearly display the coupling between average strain and average velocity, and average momentum and average stress, which is non-zero even in the symmetric case (\ref{EEffective}). The micro-structural approach directly produces the effective parameters of Eq. (\ref{EEffective}) without any \emph{a priori} assumption about the structure of the constitutive relations. These effective parameters can always be used to calculate $D^\mathrm{eff}$ and $\rho^\mathrm{eff}$, if desired.  

Fig. (\ref{FUnitCell4Layered}a) shows one unit cell of the 4-layered composite under consideration. It is composed of a heavy and stiff layer sandwiched between two layers of a light and compliant material with the whole assembly being embedded in a heavy and stiff matrix. Here we calculate the effective dynamic properties of the composite using the micro-structural method (\ref{EDeff},  \ref{Erhoeff}) with increasingly more refined discretization and compare the corresponding homogenization results with those obtained from Eqs. (\ref{EEffectiveC}, \ref{EEffectiveRho}).

\begin{figure}[htp]
\centering
\includegraphics[scale=.75]{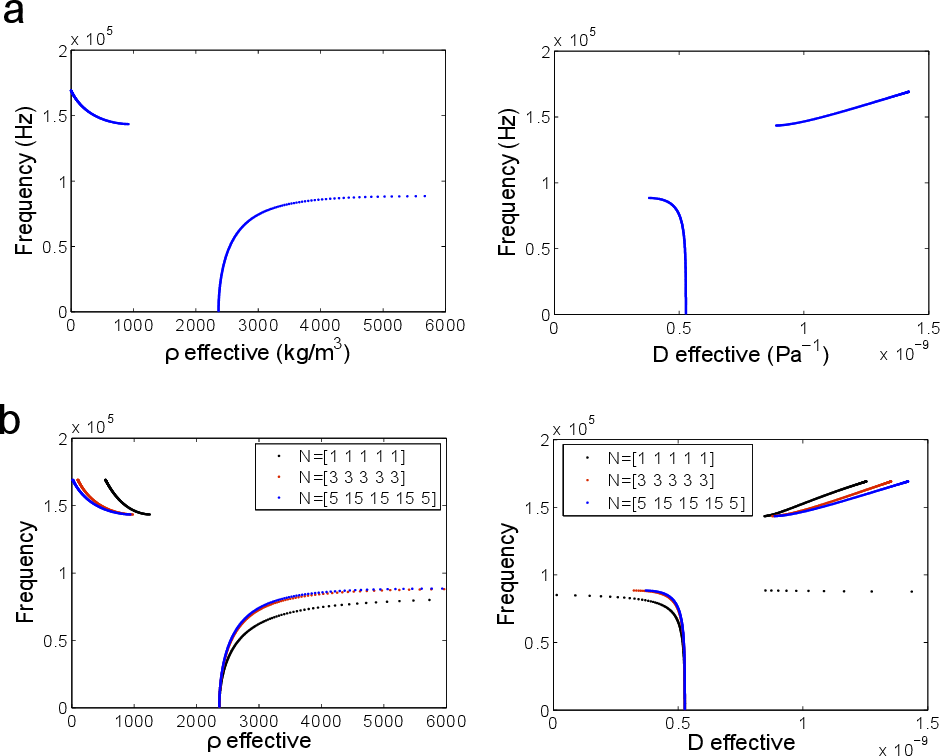}
\caption{Effective Parameters comparison: a. Homogenization results from field integral method as applied to the mixed variational formulation; b. Homogenization results from the micro-structural method with increasingly more refined discretization}\label{FEffectiveParameters4Layered}
\end{figure}

Fig. (\ref{FEffectiveParameters4Layered}a) shows the homogenization results for the first two propagating branches calculated from the integration of field variables as derived from the mixed variational formulation. Fig. (\ref{FEffectiveParameters4Layered}b) shows the homogenization results from the micro-structural formulation. This example also serves to show that the initial choice of the homogenizing parameters ($\rho_0,D_0$) is immaterial to the final homogenization results. In this example $\rho_0$ and $D_0$ are taken to be the average properties of layers 1 and 2. Therefore, the discretization of heterogeneity is applied to the entire unit cell and every layer is discretized. The effect of increasingly more refined discretization is shown in Fig. (\ref{FEffectiveParameters4Layered}b). The number of terms in the Fourier expansion ($\xi$) is kept constant at 10 for each level of discretization. From Fig. (\ref{FEffectiveParameters4Layered}b), it can be seen that when each layer is considered as just one subregion of constant eigenstress and eigenvelocity ($N=[1,1,1,1,1]$), the homogenization results deviate significantly from the results of the field integration method. As finer discretization is introduced in each layer ($N=[3,3,3,3,3]$), homogenization based on the micro-structural method tends to converge to the results of the integration method using the mixed variational formulation. For  a suitably fine discretization, especially for the case when the more compliant layer is modeled with greater number of subregions ($N=[5,15,5,15,5]$), the results of the micro-structural method approach those of the mixed formulation, with any desired accuracy.

\subsection{4-layered composite with an asymmetric unit cell}
We now present an asymmetric unit cell which serves to illustrate the nature of the coupling parameters in Eq. (\ref{EEffective}). For a symmetric unit cell, all effective parameters calculated from Eq. (\ref{EEffective}) are real-valued and yield real-valued $D^\mathrm{eff}$ and $\rho^\mathrm{eff}$. It is for the asymmetric unit cell that the coupling terms in Eq. (\ref{EEffective}) become complex-valued resulting in complex-valued $D^\mathrm{eff}$ and $\rho^\mathrm{eff}$. 

\begin{figure}[htp]
\centering
\includegraphics[scale=.75]{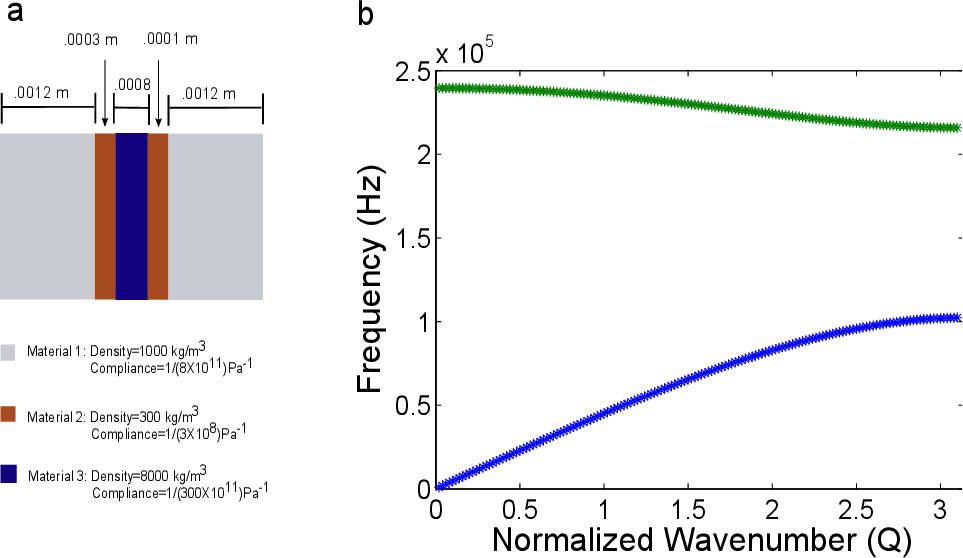}
\caption{4-layered asymmetric composite: a. Schematic of a unit cell; b. Dispersion curve for the first 2 propagating branches}\label{FUnitCell4LayeredA}
\end{figure}

Fig. (\ref{FUnitCell4LayeredA}a) shows one unit cell of the 4-layered asymmetric composite under consideration. It is composed of a heavy and stiff layer sandwiched between two equally thick layers made of different compliant materials, with this assembly being embedded in a heavy and stiff matrix. Here we calculate the effective dynamic properties of the composite using the micro-structural method (\ref{EDeff},\ref{Erhoeff}).

\begin{figure}[htp]
\centering
\includegraphics[scale=.75]{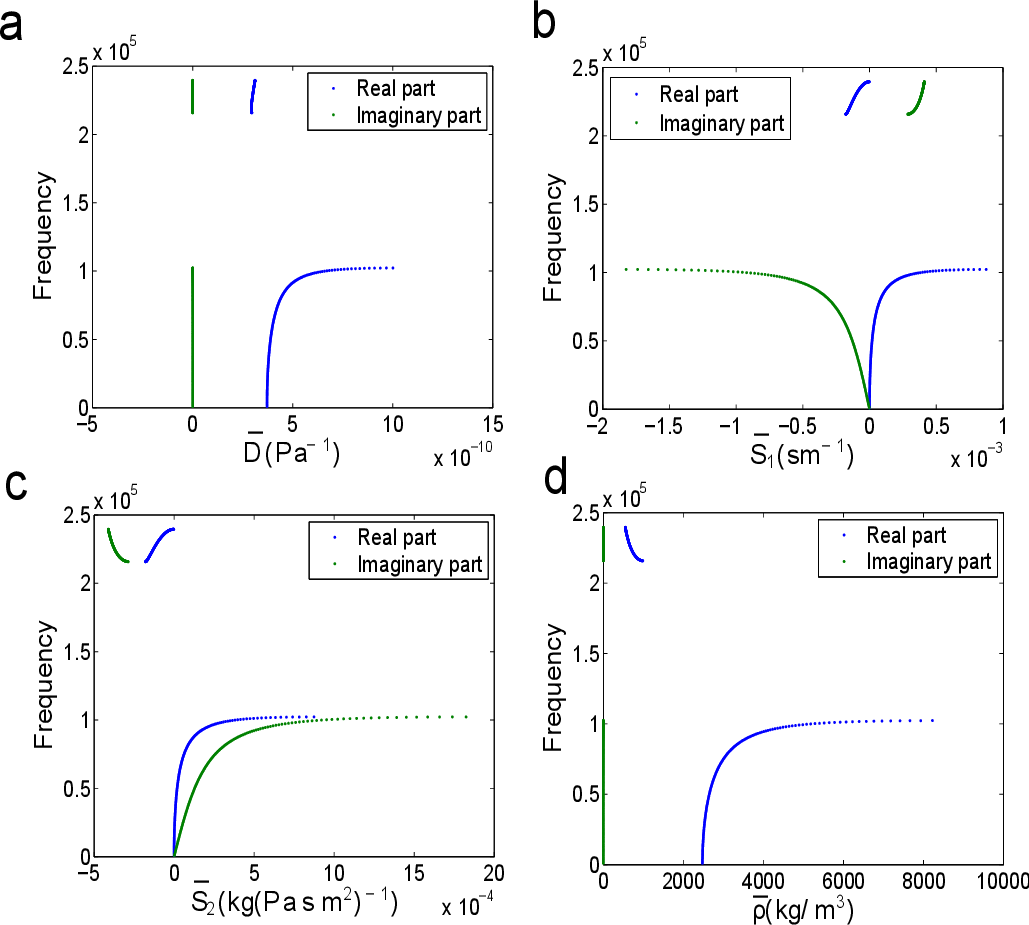}
\caption{Effective parameters calculated from the micro-structural formulation. a. $\bar{D}$, b. $\bar{S}_1$, c. $\bar{S}_2$, d. $\bar{\rho}$.}\label{FEffectiveParameters4LayeredA}
\end{figure}

Fig. (\ref{FEffectiveParameters4LayeredA}) shows the homogenization results for the first two propagating branched calculated from the micro-structural formulation (Eq. \ref{EEffective}). In this example $\rho_0$ and $D_0$ are taken to be the properties of layer 1. Therefore, the discretization of heterogeneity applies to the central three layers. The number of terms in the Fourier expansion ($\xi$) is kept constant at 10 and the discretization level of the central three layers is ($N=[15,10,15]$). The real and imaginary parts of the calculated parameters are indicated with different symbols. Figs. (\ref{FEffectiveParameters4LayeredA}a,d) clearly show that $\bar{D}$ and $\bar{\rho}$ are purely real for all propagating frequencies and Figs. (\ref{FEffectiveParameters4LayeredA}b,c) show that $\bar{S}_1$ and $\bar{S}_2$ are complex conjugates of each other.

\section{Conclusions}

A homogenization method based on micromechanical considerations \cite{SNNH,Mura} is presented. This method does not require the point-wise values of the elasto-dynamic field variables within a unit cell, instead, it produces those values to any desired degree of accuracy, if needed. Using the \emph{local} conservation and kinematical relations together with the \emph{local} constitutive relations, we have systematically deduced the overall constitutive relations for the homogenized elastic solid that depend on the frequency and wavenumber as two independent parameters.  These parameters can be related to one another in solving any special overall initial-boundary-value problems associated with the homogenized continuum.  For example, for a Bloch-form wave propagating in an infinite homogenized continuum (microstructurally layered periodic composite), one arrives at the dispersion relations that connect the wavenumber to the wave frequency, as necessary conditions for the existence of such solutions.  In general, for elastic waves in a layered unit cell, the average strain and average momentum are given in terms of the average stress and average velocity by a pair of coupled constitutive relations. We have proved that the overall compliance and mass-density in these relations are always real-valued, and that the corresponding coupling terms are always complex conjugates of each other.  For a symmetric unit cell, the coupling terms are real-valued and equal to one another. 

For illustration, we have used the overall constitutive relations to calculate the dispersion relation of a layered composite and have shown that the results converge to the exact solution for the case of a 2-layered composite. For the special problem of a Bloch-form wave propagating in an infinite homogenized solid, one may use the overall conservation and kinematical relations to define another set of effective parameters, $D^\mathrm{eff}$ and $\rho^\mathrm{eff}$, that directly relate the average strain to the average stress, and the average momentum to the average velocity, similar to the corresponding quasi-static case; see Nemat-Nasser and Hori\cite{SNNH}.  We have shown that these parameters are complex-valued for any non-symmetric unit cell.  We have presented a symmetric 4-layered example showing that these parameters as calculated from the micromechanical method converge to those calculated by the integration of field variable approach \cite{NWSA}. The microstructural method we presented here however, may be confidently used to homogenize unit cell of more complex microstructures in 2- and 3-dimensional composites where the exact dispersion relations are not known and the point-wise solution to the field equations is not available, as we shall show in a subsequent work. Finally, we presented an asymmetric 4-layered example where we have shown that the coupling terms are complex conjugates of each other and $\bar{D}$ and $\bar{\rho}$ are real-valued, whereas the corresponding $D^\mathrm{eff}$ and $\rho^\mathrm{eff}$ are complex-valued.

\section{Acknowledgement}

The authors are grateful to Professor John R. Willis and Dr. Alireza V. Amirkhizi for valuable discussions.  In particular, Dr. Amirkhizi's critical questions have impelled the authors to expand and clarify certain parts of the presentation, improving the paper. This research has been conducted at the Center of Excellence for Advanced Materials (CEAM) at the University of California, San Diego, under DARPA AFOSR Grant FA9550-09-1-0709 to the University of California, San Diego.

\appendix
\section{Homogenization by integration of field variables}\label{Appendix}

Nemat-Nasser et. al. \cite{NWSA} proposed a homogenization method of periodic elastic composites based upon the integration of field variables. Here we present a brief summary of the basic equation. For harmonic waves traveling in a layered composite with a periodic unit cell $\Omega=\{x:-a/2\leq x<a/2\}$ the field variables (displacement, velocity, stress, and momentum) take the Bloch form given by Eq. (\ref{EBloch}). The dynamic equilibrium gives,

\begin{equation}\label{EEquationOfMotion}
\nabla\hat{\sigma}+i\omega \hat{p}=0
\end{equation}

where $\nabla$ denotes differentiation with respect to $x$.

\subsection{Effective properties}

Multiply Eq. (\ref{EEquationOfMotion}) by $e^{-iqX}$ and use Eq. (\ref{EBloch}) to obtain,

\begin{equation}%\label{EStressBloch}
\nabla\left(\sigma(x)e^{iq(x-X)}\right)+i\omega p(x)e^{iq(x-X)}=0
\end{equation}

Introduce the change of variable $y=x-X$ and average with respect to $X$ over a unit cell to obtain,

\begin{equation}%\label{EStressBloch}
\nabla_y\left(\langle\sigma\rangle e^{iqy}\right)+i\omega \langle p\rangle e^{iqy}=0
\end{equation}

where

\begin{equation}\label{EAverageSigmaP}
\langle\sigma\rangle=\frac{1}{a}\int_{-a/2}^{+a/2}\sigma(x)dx;\quad \langle p\rangle=\frac{1}{a}\int_{-a/2}^{+a/2}p(x)dx
\end{equation}

Now define the \emph{mean stress} and \emph{mean momentum density} as,

\begin{equation}\label{EMeanStressMomentum}
\langle\hat{\sigma}\rangle(x)=\langle\sigma\rangle e^{iqx};\quad \langle \hat{p}\rangle(x)=\langle p\rangle e^{iqx}
\end{equation}

Observe that the mean stress and mean momentum density satisfy exactly the overall equation of motion,

\begin{equation}\label{EEquationOfMotionMean}
\nabla\langle\hat{\sigma}\rangle+i\omega\langle \hat{p}\rangle=0
\end{equation}

Define the effective mean displacement as,

\begin{equation}\label{EMeanDisplacement}
\langle \hat{u}\rangle(x)=\langle u\rangle e^{iqx};\quad \langle u\rangle=\frac{1}{a}\int_{-a/2}^{+a/2}u(x)dx
\end{equation}

Then the effective mean strain and velocity are given by,

\begin{equation}\label{EMeanStrainVelocity}
\langle \hat{\varepsilon}\rangle(x)=iq\langle \hat{u}\rangle(x);\quad \langle \hat{\dot{u}}\rangle(x)=-i\omega\langle \hat{u}\rangle(x)
\end{equation}

Finally, the averaged stress-strain relation becomes, 

\begin{equation}%\label{EMeanStrainVelocity}
D^\mathrm{eff}\langle \hat{\sigma}\rangle(x)=\langle \hat{\varepsilon}\rangle(x)
\end{equation}

Based on Eq. (\ref{EMeanStressMomentum}, \ref{EMeanStrainVelocity}), $D^\mathrm{eff}$ is given by,

\begin{equation}\label{EEffectiveC}
D^\mathrm{eff}=\frac{iq\langle u\rangle}{\langle \sigma\rangle}
\end{equation}

Similarly, the effective density ($\rho^\mathrm{eff}$) is given by,

\begin{equation}\label{EEffectiveRho}
\rho^\mathrm{eff}=\frac{\langle p\rangle}{-i\omega\langle u\rangle}
\end{equation}

Since $\langle\hat{\sigma}\rangle$ and $\langle \hat{p}\rangle$ satisfy the equation of motion (Eq. \ref{EEquationOfMotionMean}), it is seen that the effective material parameters defined above automatically satisfy the dispersion relation,

\begin{equation}%\label{EEffectiveRho}
\frac{1}{D^\mathrm{eff}\rho^\mathrm{eff}}=\frac{\omega^2}{q^2}
\end{equation}

\end{document}